\documentclass[twocolumn]{aastex63}
\usepackage{amsmath}
\usepackage{float}
\usepackage{enumerate}
\usepackage{soul}
\usepackage{xcolor}
\usepackage{graphicx}
\usepackage{amssymb}
\usepackage{hyperref}
\usepackage{listings}
\usepackage{orcidlink}
\usepackage{tikz}

\submitjournal{ApJ}
\shorttitle{Polar Crown Filament formation via supergranulation}
\shortauthors{Chen \& Xia}
\begin{document}
\soulregister{\cite}{7}
\soulregister{\ref}{7}
\soulregister{\citep}{7}
\soulregister{\citet}{7}

\title{Formation of Polar Crown Filaments Magnetic Fields by Supergranular Helicity Injection}

\correspondingauthor{Chun Xia}
\email{chun.xia@ynu.edu.cn}

\author{Huanxin Chen}
\affiliation{School of Physics and Astronomy, Yunnan University, \\
 Kunming 650500, China}

\author{Chun Xia}
\affiliation{School of Physics and Astronomy, Yunnan University, \\ 
 Kunming 650500, China}
\affiliation{Yunnan Key Laboratory of Solar Physics and Space Science, 650216, China}
 \affiliation{National Astronomical Observatories, Chinese Academy of Sciences, \\
 Beijing 100101, China}

\author{Hechao Chen}
\affiliation{School of Physics and Astronomy, Yunnan University, \\
 Kunming 650500, China}

\begin{abstract}
To understand the magnetic fields of the polar crown filaments (PCFs) at high latitudes near polar regions of the Sun, we perform magnetofrictional numerical simulations on the long-term magnetic evolution of bipolar fields with roughly east--west polarity inversion lines (PILs) in a three-dimensional (3D) spherical wedge domain near polar regions. The Coriolis effect induced vortical motions at the boundaries of several supergranular cells inject magnetic helicity from the photospheric boundary into the solar atmosphere. Supergranular-scale helicity injection, transfer, and condensation produce strongly sheared magnetic fields. Magnetic reconnections at footpoints of the sheared fields produce magnetic flux ropes (MFRs) with helicity signs consistent with the observed Hemispheric Helicity Rule (HHR). The cross-sectional area of MFRs exhibits an uneven distribution, resembling a "foot-node-foot" periodic configuration. Experiments with different tilt directions of PILs indicate that the PCFs preferably form along PILs with the western end close to the polar region. The bending of PILs caused by supergranular flows, forming S-shape (Z-shape) PIL segments, promotes the formation of dextral (sinistral) MFRs. The realistic magnetic models we got can serve as starting points for the study of the plasma formation and eruption of PCFs.
\end{abstract}


\section{Introduction} \label{sec:intro}

Solar prominences/filaments are relatively cool and heavy plasma suspended in the million-degree corona by special magnetic field structures. Solar filaments are widespread across the entire latitude of the Sun. Polar crown filaments (PCFs) are quiescent filaments appearing at high latitudes close to the polar regions. PCFs form in filament channels above the polarity inversion lines (PILs) between the polar region magnetic fluxes and the opposite-polarity fluxes of dead active regions transported from lower latitudes. Magnetic fields with a dominant horizontal component, nearly parallel to the PIL in a filament channel, support the filament material against gravity \citep{Martin_1998,Mackay_2010}. The overall magnetic topology of the filament channels of PCFs is likely to be a helical magnetic flux rope (MFR), which is supported by many observational evidences and magnetohydrodynamic (MHD) simulations \citep{Bommier1994,Wang2010,Gibson2010,Bak2013,Xia_2014,Xia_2016,Fan2018}. To explain the strongly sheared magnetic field and the origin of MFRs in filament channels, the flux cancellation models \citep{Van_1989,Xia2014} rely on photospheric shearing and converging motions along the PILs and the flux emergence models \citep{Fan_2001,Manchester_2001} depend on the inheritance of a pre-existing MFR under the photosphere during flux emergence. However, PCFs form in quiescent regions where no significant flux emergence has been observed \citep{Mackay2008}. Observations have not found large-scale shearing and converging flows along the PILs of quiescent filament channels where ordinary supergranular flows prevail \citep{Rondi2007,Schmieder2014}. 

\citet{Antiochos_2013} proposed the magnetic helicity condensation theory to explain the sheared magnetic field in filament channels. In his theory, magnetic helicity is thought to be injected into the upper atmosphere through small-scale photospheric vortical motions between convective (super)granular cells \citep{Duvall2000,Langfellner2015}. Adjacent magnetic flux tubes with the same sense of twist (helicity) may experience component magnetic reconnection at the contact points between the tubes, thus the two tubes merge into one, and the twist (helicity) of the tubes transfers to the periphery. This process occurs continuously from small-scale to large-scale flux tubes, leading to the cascade of injected helicity from the small-scale to the periphery of the largest-scale flux tube, e.g., around the PIL, where the magnetic helicity condenses. The ideal MHD simulations \citep{Zhao_2015,Knizhnik_2015,Knizhnik2017} supported the theory and demonstrated the accumulation of magnetic shear at PILs. These models neglected the primary diverging motions of convective cells on the photosphere and could not produce MFRs. Recently, \citet{Liu2022} included time-dependent diverging motions of supergranular cells on the photospheric boundary and simulated the formation of a negative-helicity MFR in the northern hemisphere via magnetofrictional evolution. The converging motions at the boundaries of supergranules became vortical motions under the effect of Coriolis force and injected negative magnetic helicity into the flux tubes. The helicity condensation generates PIL-aligned magnetic sheared arcades, which are then reconnected in a head-to-tail fashion under the converging flows between supergranules at various sites of the PIL to form an MFR.

Regarding the chirality, filaments and filament channels are classified into dextral type and sinistral type. When an observer stands on the positive-polarity side of the PIL and looks toward the PIL, a dextral (sinistral) filament has the magnetic field component parallel to the PIL, pointing to the right (left) \citep{Martin_1994}, coexisting with left-skewed (right-skewed) overlying sheared magnetic arcades, whose parallel-to-the-PIL magnetic field components point to the same direction as the axial field of the filament does \citep{Martin_1998,Chen2014}. The dextral (sinistral) filament channels contain magnetic fields with negative (positive) magnetic helicity. Statistical studies found that the chirality of filaments, especially the quiescent filaments have a strong hemispheric pattern, such that dextral filaments mostly appear in the northern hemisphere, while sinistral filaments in the southern hemisphere \citep{Martin_1994,Pevtsov_2003,Ouyang2017}. More than 80\% quiescent filaments and 76\% active-region filaments followed the hemispheric pattern with their chirality deduced from the bearing sense of filament barbs \citep{Martin_1994,Pevtsov_2003}, and the percentage increased to 93\% (83\%) for quiescent (active-region) filaments with the chirality found by the skew of the drainage sites of erupting filaments \citep{Ouyang2017}. Generally speaking, the filament magnetic fields follow the hemispheric helicity rule (HHR)/Hemispherical Chirality Pattern, i.e., the negative (positive) magnetic helicity dominates in the northern (southern) hemisphere, which is independent of solar cycles and also found in the magnetic fields of active regions, coronal sigmoids, and interplanetary magnetic clouds \citep{Rust1994,Pevtsov_1995,Zirker_1997}. 

Early explanations of the HHR considered the role of large-scale velocity fields in magnetic helicity injection. The differential rotation was thought to inject the correct-sign helicity into the emerged bipolar fields of the north--south PILs \citep{DeVore_2000}. \citet{Rust1994} conjectured that the subsurface equatorial jet stream could twist subsurface magnetic flux tubes before their emergence, injecting negative (positive) helicity into the northern (southern) part. However, observations on high-latitude quiescent filaments, which have predominantly east--west PILs, found the direction of axial fields opposite to the hypothetical ones produced by differential rotation acting on bipolar coronal magnetic fields \citep{Rust1967,Leroy1983}. A series of magnetofrictional simulations on the HHR of filaments, starting from idealized magnetic fields \citep{Mackay2001,Mackay2005} or observational-data-based magnetic fields \citep{vanBallegooijen1998,Yeates2007,Yeates2008}, considered the magnetic flux transport effects of differential rotation, meridional flow, and magnetic diffusion on the photosphere. In these models, the magnetic fields of filaments largely inherit the magnetic helicity of the initial or emerged bipolar active regions. Combined with negative (positive) helicity bipoles emerging in the northern (southern) hemisphere, these models obtained over 90\% agreement with the observed chirality of the filaments, especially for filaments below $60^\circ$ latitude. \citet{Yeates_2012} extended these models to the global magnetic field evolution covering the entire solar cycle 23. They reproduced the hemispheric pattern during the rising phase of the solar cycle. However, during the declining phase, their model presented dominating sinistral (dextral) chirality of high-latitude filaments in the northern (southern) hemisphere, contradicting the observations. The main reason is that the differential rotation acting on the east--west PILs, typical for PCFs between the polar and active-latitude fluxes inject positive (negative) magnetic helicity in the northern (southern) hemisphere, in conflict with the hemispheric rule \citep{vanBallegooijen1998}.

Observations \citep{Duvall2000,Langfellner2015,Requerey2018} have found that, in the northern (southern) hemisphere, the diverging flows of supergranular cells present clockwise (anticlockwise) vortical motions leading to anticlockwise (clockwise) vortical motions in converging flows around the junction points of several supergranules. These phenomena are caused by the Coriolis force \citep{Hathaway1982, Egorov2004}. In each hemisphere, the sign of the magnetic helicity injected by these vortical motions between supergranules, where magnetic fluxes accumulate, is consistent with the HHR. Since the helicity condensation theory and models have proven the important role of supergranular helicity injection in the formation of filament channels, a natural conjecture is that the hemispheric pattern of PCFs may be determined by the helicity injection from supergranulations under Coriolis force. \citet{Mackay_2018} used a temporally and spatially averaged statistical approximation of helicity injection--condensation and numerically reproduced the hemispheric pattern of high-latitude filaments in both the rising and declining phases of a solar cycle with global magnetofrictional simulations. Their global-scale models could not resolve supergranular scale motions and rely on statistically averaged large-scale rotations to inject helicity. Recently, \citet{Liu2022} used numerically resolved supergranular flows as a boundary-driven condition and successfully produced an MFR for middle-latitude quiescent filaments starting from a north--south PIL. In this work, we use their model to investigate the origin of the hemispheric pattern of PCFs starting from east--west PILs.

\section{Numerical Methods} \label{sec:methods}

Our computational domain is a partial spherical shell of the range $1 R_{\odot}<r<1.7 R_{\odot}, 37^\circ<\theta<73^\circ, 0^\circ<\phi<90^\circ$. The domain is discretized into a four-level adaptive-mesh-refinement (AMR) spherical mesh with $512\times256\times256$ effective resolution and the logarithmic stretching in $r$ \citep{Xia_2018}. The initial magnetogram $B_r=\sum_{i=1}^2 B(\theta,\phi)_i$ on the photosphere is a superposition of two quasi-elliptic Gaussian-like flux regions:

\begin{equation}
B(\theta,\phi)_i = 
\begin{cases} 
 (-1)^i B_0 e^{-(\frac{\theta - \theta_i}{S_i})^2}, & \phi_{ei} < \phi < \phi_{wi} \\
 (-1)^i B_0 e^{-(\frac{\theta - \theta_i}{S_i})^2-(\frac{\phi - \phi_{i}}{P_i})^2}, & 
    \phi \le \phi_{ei}~{\rm or}~\phi \ge \phi_{wi} , (i = 1\ or\ 2)
\end{cases}
\label{eq:magnetogram}
\end{equation}

where $B_0=20$ G, $\theta$ coordinates of the center of the flux regions $\theta_{1}=60.5^\circ, \theta_{2}=50.5^\circ$, decay scales in the theta direction $S_1=6.3^\circ, S_2=7.2^\circ$, decay scales in the phi direction $P_1=14.4^\circ, P_2=13.9^\circ$, decay-starting $\phi$ coordinates at the east and west end $\phi_{e1}=37^\circ, \phi_{w1}=51^\circ, \phi_{e2}=39^\circ, \phi_{w2}=53^\circ$. Two opposite fluxes (Figure~\ref{fig:NSMFR} (a)) symmetric about the latitude line correspond to subscripts 1 and 2 in Equation~\ref{eq:magnetogram}. This bipolar flux distribution has an east--west PIL at latitude $55^\circ$. Using functions in the PDFI\_SS software \citep{Fisher_2020}, the initial potential magnetic field is extrapolated from the magnetogram on the radially stretched spherical mesh, with periodic $\phi$ boundaries, perfectly conducting $\theta$ boundaries (zero normal magnetic flux), and the source surface at $1.7 R_{\odot}$. The magneto-friction equations \citep{Guo_2016} without explicit resistivity are solved using the MPI-AMRVAC software \citep{Xia_2018}. We use periodic boundary conditions on the $\phi$ boundaries, perfect conducting $\theta$ boundary condition with zero tangential electric fields, third-order equal normal gradient and second-order zero normal gradient extrapolation of magnetic field for the photospheric boundary and the outer $r$ boundary, respectively, and zero velocity condition for boundaries other than the photospheric boundary. The numerical schemes and the photospheric-boundary supergranular velocity fields based on Voronoi tessellation are detailed in the paper of \citet{Liu2022} (In the Voronoi tessellation, 2D Poisson randomly distributed seed points serve as the centers of supergranular cells, and every point in a supergranular cell is closer to the seed point of the cell than any other seed point.). The differential rotation velocities are added with zero value at the middle $\theta=55^\circ$. The latitude-dependence of Coriolis force is approximated by adding $\cos \theta$ factor to the vortical velocities of supergranules. All boundary-driven velocities are speeded up by a factor of 5 to save computational time. The time unit, with the speed up considered, is 5 hr. The unit of magnetic field is 2 G.

As listed in Table~\ref{tab:model}, we simulated ten models. The domain setup described above is for the NS (North Standard) model centered at latitude $55^\circ$ in the northern hemisphere. The SS (South Standard) model is mirror-symmetric to the model NS about the equatorial plane with anticlockwise rotating supergranular cells in the southern hemisphere. These two models serve as standard cases for each hemisphere. The model SC relative to NS in different hemispheres during the same solar cycle. In the models, NW1 and NW2, the vortical velocities of supergranules are weakened to the $1/3$ and $2/3$ of the standard case, respectively. The model NC has zero vortical speed excluding the Coriolis effect. Observations found that many PCFs are not exactly along the east--west direction. \cite{Tlatov2016} counted H$\alpha$ filaments from 1919 to 2014, and found that the tilt angles of filaments concerning the solar equator have a hemispheric rule, such that the southeast--northwest (northeast--southwest) oriented filaments dominate in the polar regions of northern (southern) hemisphere. To simulate such cases, we modify the initial magnetogram to make the PIL tilted with an angle of $20^\circ$ away from the east--west direction in models NT1, NT2, ST1, and ST2. The experiments with angles of $5^\circ$, $10^\circ$, and $30^\circ$ have also been completed, and their results show little difference. As the tilt angle increases, there are slight variations, but the trend lacks a definable pattern. We present the case at $20^\circ$ as representative of the mean value of observations \citep{Tlatov2016}.

\begin{table*}
   \begin{center}
   \caption{A list of models with different setups}
   \label{tab:model}
   \begin{tabular}{lcccc}
   \hline
   Name & Hemisphere & Coriolis effect & PIL direction & Chirality / Helicity\\ 
   \hline
	    NS  & north & 1& east--west & dextral / negative\\
	    NW2 & north & $2/3$& east--west & dextral / negative\\
	    NW1 & north & $1/3$& east--west & dextral / negative\\
	    NC  & north & 0& east--west & dextral / negative\\
	    NT1 & north & 1& southeast--northwest & dextral / negative\\
	    NT2 & north & 1& northeast--southwest & dextral / negative\\
	    SS (SC) & south & 1& east--west & sinistral / positive\\
	    ST1 & south & 1& northeast--southwest & sinistral / positive\\
	    ST2 & south & 1& southeast--northwest & sinistral / positive\\
  \hline
  \end{tabular}
  \\
  \footnotesize{Notes: models NW1 and NW2 have weaker vortical speeds of supergranulations than model NS, and model NC has zero vortical speed excluding the Coriolis effect. Model SC is southern model of the same cycle for model NS. Models NT1, NT2, ST1, and ST2 have a tilted PIL relative to models NS and SS.}
  \end{center}
\end{table*}


\section{Formation of flux rope and HHR/hemispherical chirality pattern} 
\begin{figure*}[htbp]
\centering
\includegraphics[width=\textwidth]{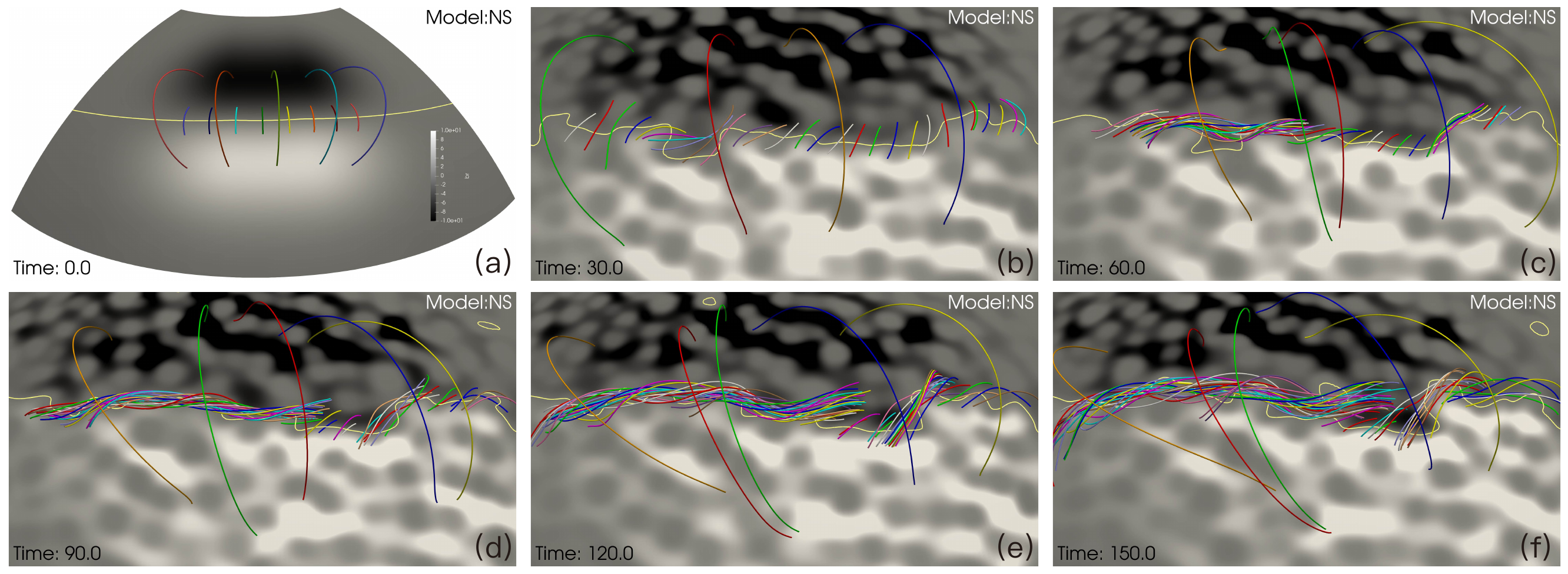}
\caption{Time evolution of the standard model NS at time 0 (a), 30 (b), 60 (c), 90 (d), 120 (e), and 150 (f). The photosphere is colored by the radial magnetic field with white positive flux and black negative flux saturated at $\pm$ 20 G (The values in the colorbar are dimensionless). The yellow line on the photosphere is the PIL. Magnetic field lines in different colors are plotted through uniformly sampled points along the PILs on horizontal spherical slices at 1.0035, 1.012, and 1.09 solar radii.}
\label{fig:NSMFR}
\end{figure*}

The time evolution of the model NS is shown in Figure~\ref{fig:NSMFR}. Initially (Figure~\ref{fig:NSMFR} (a)), the photospheric magnetogram colored by the radial magnetic field presents a smooth dipole field and the PIL is quite straight in the east--west direction. The high and low magnetic arcades are in potential states. At time 30 (Figure~\ref{fig:NSMFR} (b)), the photospheric magnetic fluxes are swept by the diverging supergranular motions to boundaries of supergranules forming a magnetic network. The low and short magnetic arcades are skewed clockwise with negative magnetic helicity because the helicity condensation process accumulates negative helicity near the PIL, which overcomes the positive helicity injection by differential rotation. The high and long magnetic arcades become slightly skewed anticlockwise with positive helicity injected by differential rotation. A detail worth noticing is that the PIL is deformed and some bending PIL segments are generated, i.e., the initially straight PIL becomes irregular. At time 60 (Figure~\ref{fig:NSMFR} (c)), a short MFR is formed above the eastern part of the whole PIL and its chirality is dextral. Strongly sheared magnetic arcades (SMA) are found near two ends of the MFR, which are later annexed into the MFR. At time 90 (Figure~\ref{fig:NSMFR} (d)), a longer MFR spanning half of the PIL region can be found, resulting from the extension of the original short flux rope in the eastern region due to the accumulation of negative helicity and magnetic reconnection on the photosphere. By the time 120 (Figure~\ref{fig:NSMFR} (e)), the axial magnetic field lines of the MFR form a weak sigmoidal shape with a low-lying middle section and two rising ends. The low magnetic arcades in the western area are further sheared above S-shape bending PIL segments. The high overlying magnetic arcades are clearly sinistral with positive helicity, opposite to the helicity of the MFR. By the time 150 (Figure~\ref{fig:NSMFR} (f)), the formed MFR continues to grow, and it can be expected that the short magnetic arcades in the western area will eventually be integrated into the MFR. After the formation of a complete flux rope, it gradually elevates and expands over a long period. We can find magnetic dip structures like feet connected to the photosphere in the MFR. Subsequently, an eruption or a sustained stable state ensues, the development of which depends on the configuration of the magnetic field and the evolution time. Our standard model NS can reproduce an MFR, the magnetic topology of a high-latitude PCF. The footpoints of the MFR are almost rooted at the boundary of the supergranules, where the magnetic flux is very strong. The negative magnetic helicity produced by the small-scale vortices at supergranular boundaries propagates and condenses to the PIL, which can counteract and dominate the helicity of the wrong sign injected from the differential rotation.

\begin{figure*}[htbp]
\centering
\includegraphics[width=\textwidth]{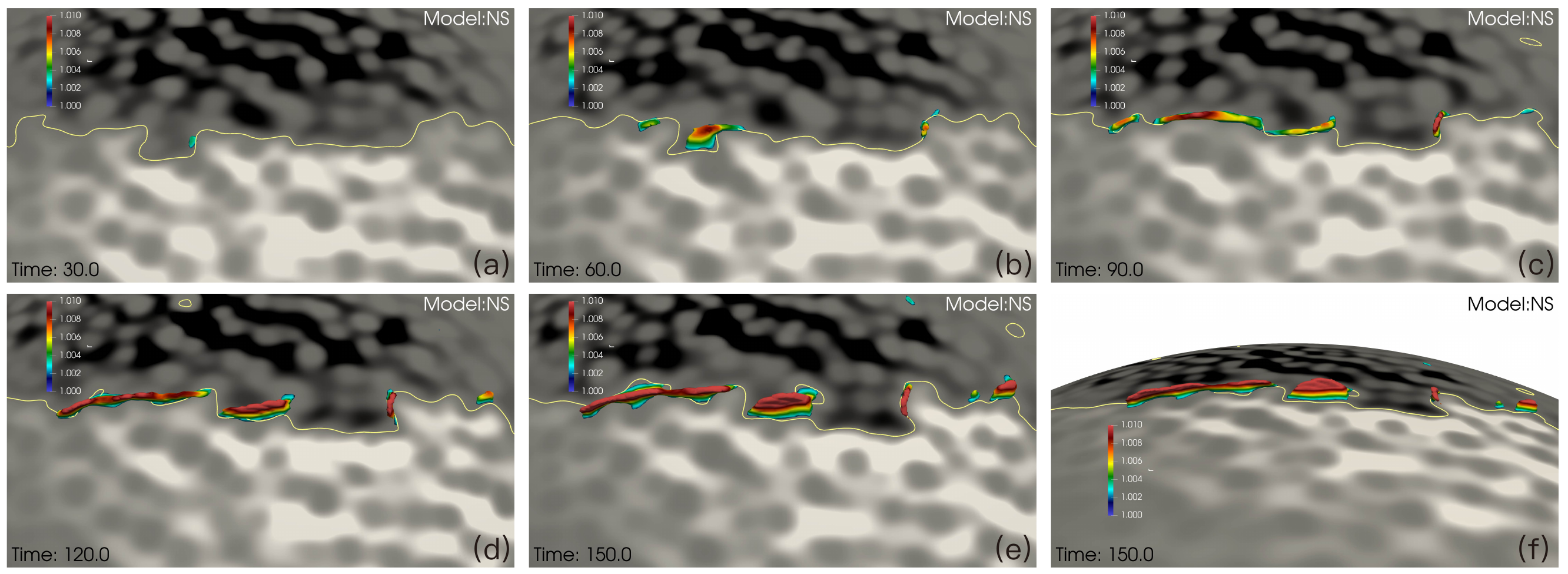}
\caption{Time evolution of the magnetic dip regions in the standard model NS at time 30 (a), 60 (b), 90 (c), 120 (d), and 150 (e, f). Cells in dip regions have a positive radial component of the curvature of the magnetic field and less than 10\% proportion of the radial component of the magnetic field. The magnetic dip regions are shown as isosurfaces. The colors of the dip regions indicate heights, saturated in red above $1.01R_{\odot}$. Panels (e) and (f) are the same data under different viewing angles. The rest of the features are explained in Figure~\ref{fig:NSMFR}. An animation of the evolution of the magnetic dips in model NS is available. The animation proceeds from time 0 to 150. Small magnetic dips (representing small flux ropes) are formed first, followed by gradual connections. The duration of the animation is 10 seconds.}
\label{fig:NSdip}
\end{figure*}

Magnetic dips are concave upward magnetic structures that can collect and support the dense prominence plasma against gravity. They therefore can roughly represent prominence plasma structures. Figure~\ref{fig:NSdip} shows magnetic dips as isosurfaces of magnetic dip criterion ($(\mathbf{b}\cdot\nabla\mathbf{b})_r>0$ and $B_r/B<0.1$, $\mathbf{b}=\mathbf{B}/B$) at the corresponding moments as Figure~\ref{fig:NSMFR}. At time 30 (Figure~\ref{fig:NSdip} (a)), the first dip region appears as a low-lying lump above a bending PIL segment. At time 60 (Figure~\ref{fig:NSdip} (b)), the first lump of dips becomes larger and higher, and multiple low-lying lumps appear along the PIL. Reaching time 90 (Figure~\ref{fig:NSdip} (c)), there are elongated dip regions, which resemble the shape of filaments. As time passes, the filamentary dip regions continue to grow higher. Until time 150 (Figure~\ref{fig:NSdip} (e) (f)), it can be seen that several foot-like structures in cyan forms. The foot-like structures are the lowest parts of an MFR because they are threaded by helical magnetic field lines of the MFR. Between these structures, there are gaps like bridge culverts. The distribution of the ``bridge culverts" and ``foot" structures are discussed in detail in the model NT1 below. The evolution of the magnetic dips predicts that the dips in growing MFRs are sufficient to support the forming of prominence plasma in a realistic shape. However, the foot-like structures of dips in the model are narrow at the top and wide at the bottom in the radial direction, which differs from the observed feet connected to the photosphere, probably because there are not real feet structures but transient structures or threads set that appear first. The present simulation is only a magnetofrictional model in the early stage of MFR formation, and we are only interested in the magnetic structure, expecting to find more reasonable explanations for the feet structure in future simulations, where plasma is added or the MFR is more mature. 

\begin{figure*}[htbp]
\centering
\includegraphics[width=\textwidth]{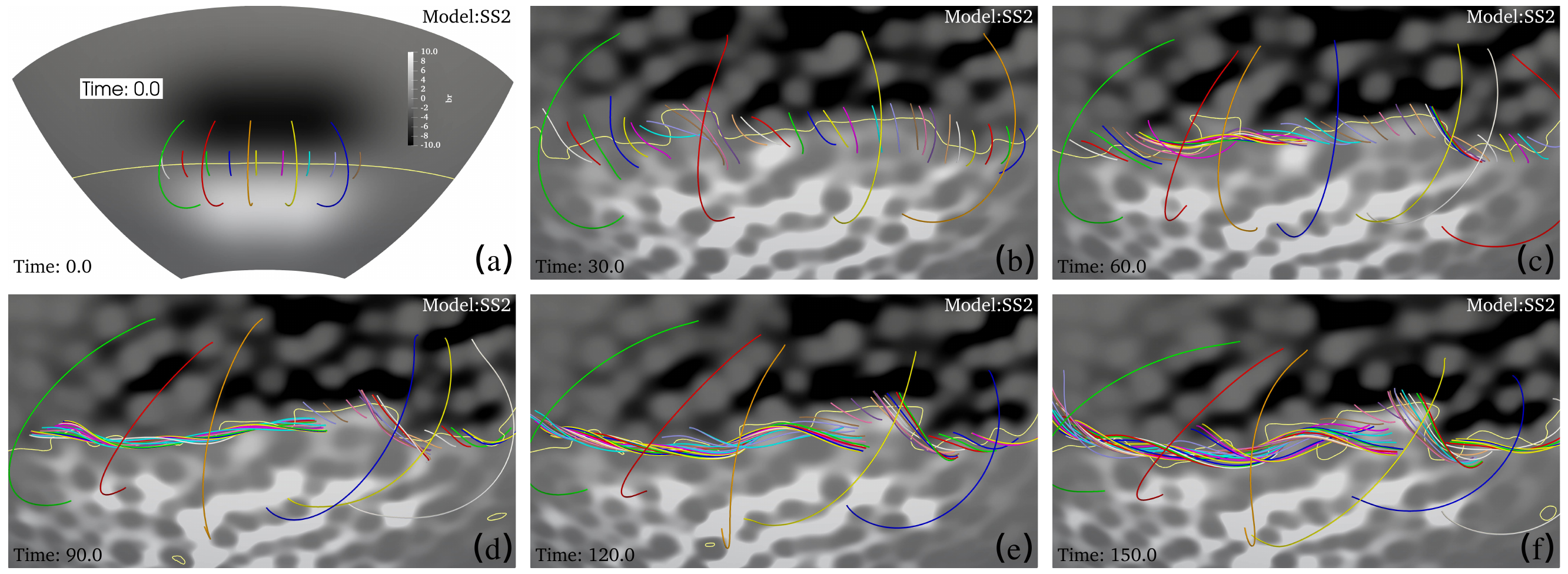}
\caption{The image of magnetic field lines in the model SC at time 0 (a), 30 (b), 60 (c), 90 (d), 120 (e), 150 (f). The features on the image are similar to Figure~\ref{fig:NSMFR}.}
\label{fig:SCMFR}
\end{figure*}

\begin{figure*}[htbp]
\centering
\includegraphics[width=\textwidth]{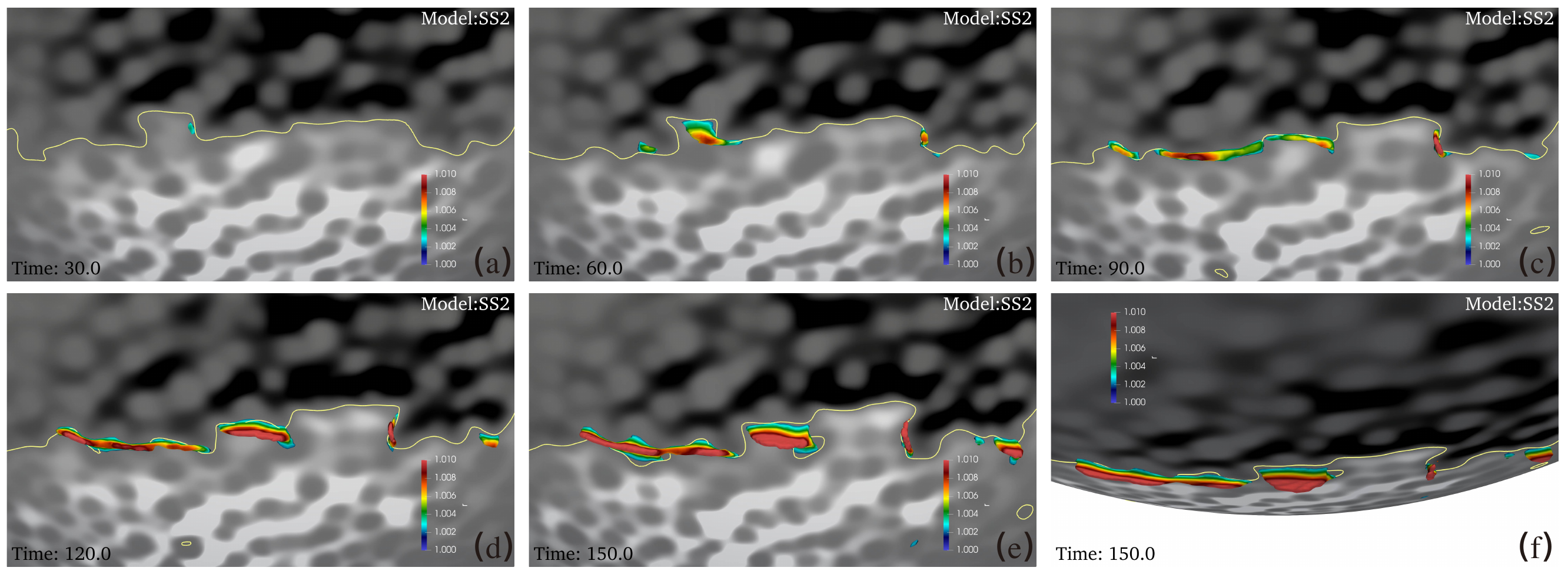}
\caption{The image of magnetic dips in model SC at time 30 (a), 60 (b), 90 (c), 120 (d), 150 (e, f). The features on the image are similar to Figure~\ref{fig:NSdip}.}
\label{fig:SCdip}
\end{figure*}

To fully verify the consistency of our models with the HHR, we established the model SC in the southern hemisphere as an experiment within the same solar cycle as the model NS. Figure~\ref{fig:SCMFR} is similar to the evolution image of Figure~\ref{fig:NSMFR} but in the southern hemisphere. The evolution image of dips in the model SC is also in the same solar cycle as the model NS (Figure~\ref{fig:SCdip}). It is interesting to note that the MFRs in the southern hemisphere have sinistral chirality because of the condensation of positive magnetic helicity by supergranular vortical motions, and the high magnetic arcades have dextral chirality because of the differential rotation. The MFRs are negative/positive helicity or dextral/sinistral chirality in the northern/southern hemisphere, which is consistent with the HHR.

\begin{figure*}[htbp]
\centering
\includegraphics[width=\textwidth]{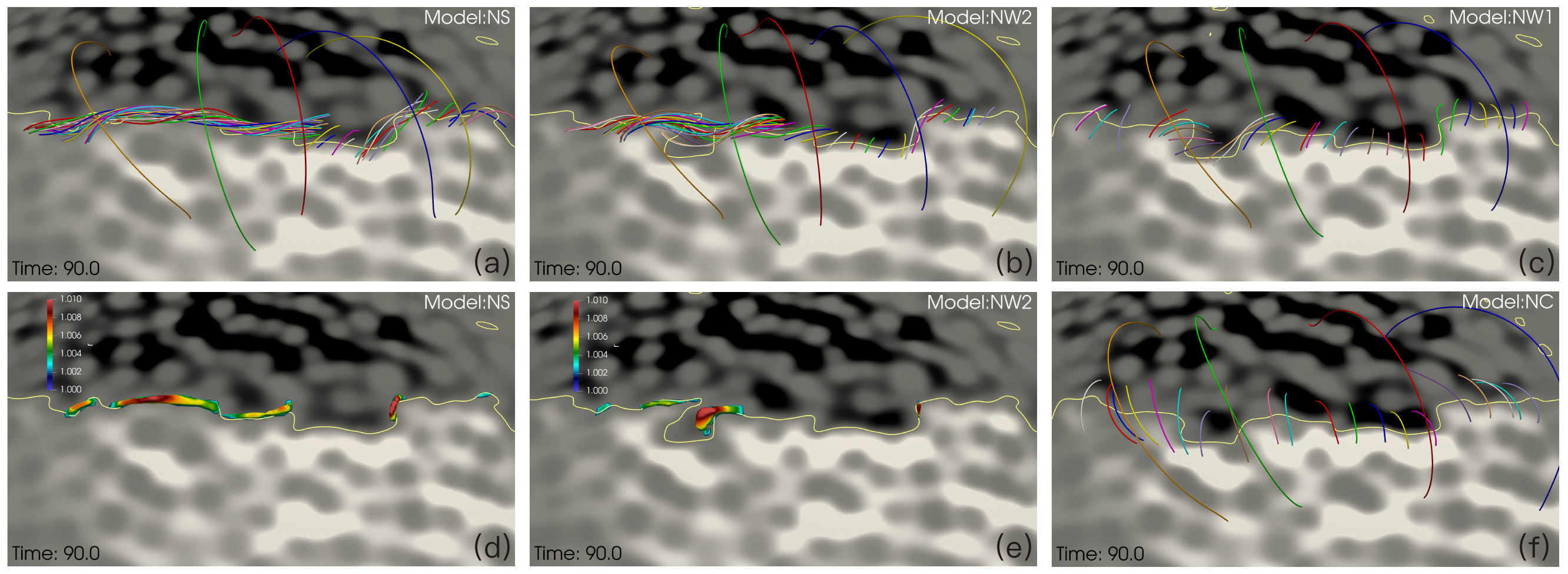}
\caption{Comparison of magnetic structures between models NS (a, d), NW2 (b, e), NW1 (c), and NC (f) at time 90. The vortical speed of supergranules in models NW2 and NW1 are 2/3 and 1/3 of those in model NS, respectively, while model NC (f) has zero vortical speed. The magnetic dips of model NS (d) and model NW2 (e) are presented in the same way as Figure~\ref{fig:NSdip}.}
\label{fig:NW}
\end{figure*}

\citet{Liu2022} found that the effect of Coriolis force plays a key role in the helicity injection and condensation for quiescent filaments. We constructed models NC, NW1, and NW2 in the northern hemisphere to explore different rates of helicity injection under different vortical speeds of supergranules. The parameters of the vortical speed for these models are listed in Table~\ref{tab:model}. Figure~\ref{fig:NW} is a comparative image of the four models at time 90. The standard model NS successfully forms a typical MFR structure above the PIL (Figure~\ref{fig:NW} (a)). In contrast, when the vortical speed is reduced to 2/3 in the model NW2 (Figure~\ref{fig:NW} (b)), the shearing of the low-lying magnetic field lines is weaker and a hybrid magnetic structure appears above the bending PIL section with a weakly twisted MFR and surrounding SMAs. As the evolution continues, MFRs can form along the PIL in model NW2. When the vortical speed is reduced to 1/3 of that in the standard model (Figure~\ref{fig:NW} (c)), even though no strongly sheared magnetic arcades and MFRs are found, the magnetic field lines near the PIL are weakly sheared with negative helicity because the negative helicity condensation overcomes the positivity helicity from the differential rotation. However, as the evolution continues, MFRs do not form. In the model NC (Figure~\ref{fig:NW} (f)), there are no small-scale vortical motions to inject magnetic helicity. The low-lying magnetic field lines are weakly sheared with positive magnetic helicity by the differential rotation, contradicting the HHR. Comparing model NS with NW2, the size of the dip regions and the cross-section of MFRs is positively correlated to the vortical speed at a given time (Figure~\ref{fig:NW} (d) (e)). The dip regions of model NW2 at time 90 have a similar size to the ones of model NS at time 60. It indicates that the MFRs form later when the vortical speed is smaller. The model NW1 and NC do not have any dip region at all times. We continue to run the model NW1 until time 350, yet we do not find any dip region or strongly sheared arcades. The interaction between the magnetic field lines seems to remain relatively calm when the vortical speed is reduced to a certain level. Even if there is a continuous injection, the negative magnetic helicity is not able to accumulate around the PIL. Therefore, the vortical speed of the supergranules significantly influences the formation speed of MFRs.

\section{Direction of PIL}

\begin{figure*}[htbp]
\centering
\includegraphics[width=\textwidth]{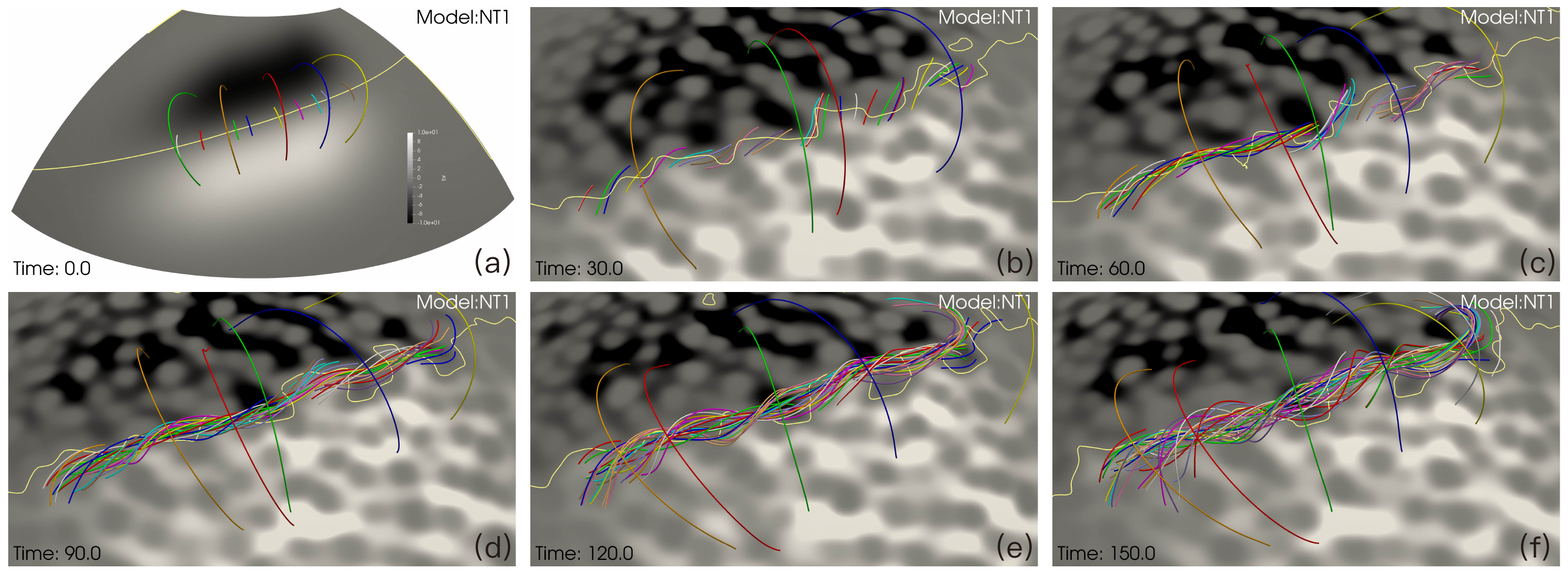}
\caption{Time evolution of the magnetic field of the model NT1 at time 0 (a), 30 (b), 60 (c), 90 (d), 120 (e), 150 (f). The features on the image are similar to Figure~\ref{fig:NSMFR}. The PIL has an angle of $20^\circ$ with the west end closer to the pole.}

\label{fig:NT1MFR}
\end{figure*}

We established models NT1, NT2, ST1, and ST2 with different PIL tilt angles relative to the east--west direction, to investigate whether the direction of PIL has any effect on the formation of MFRs in polar regions. The PIL directions of these models are listed in Table~\ref{tab:model}. As shown in Figure~\ref{fig:NT1MFR} (a), the PIL of model NT1 appears in the southeast--northwest direction. At time 30 (Figure~\ref{fig:NT1MFR} (b)), the short low-lying field lines near the distorted PIL present as strongly sheared loops with larger skew angles than the ones in model NS at the same time (Figure~\ref{fig:NSMFR} (b)). The higher magnetic arcades have a weakly dextral shearing or nearly potential state in contrary to the model NS with sinistral high arcades, due to the similarly anticlockwise rotating speed of arcades and the tilted PIL caused by differential rotation flows. At time 60 (Figure~\ref{fig:NT1MFR} (c)), there is a long MFR in the southeast part of PIL, a short MFR in the northwest, and sheared arcades in between. At time 90 (Figure~\ref{fig:NT1MFR} (d)), some magnetic field lines in green have been connected to become the axial part of a large MFR, spanning from the southeast to the northwest of the PIL, while twisted field lines in the outer layer of the MFR are still disconnected with footpoints near the middle part of the PIL. By the time 120 (Figure~\ref{fig:NT1MFR} (e)), twisted field lines in the outer layer of the MFR are completely connected. At time 150 (Figure~\ref{fig:NT1MFR} (f)), the MFR grows fatter and higher with footpoints mainly located in the southeast and northwest ends. It can be seen that the MFRs in NT1 form earlier than those in model NS, and the MFRs of NT1 are larger and more mature at the same moments. Therefore, the formation of the long MFR from a series of small MFRs in model NT1 is easier and more successful due to the southeast--northwest tilt of the PIL.

\begin{figure*}[htbp]
\centering
\includegraphics[width=\textwidth]{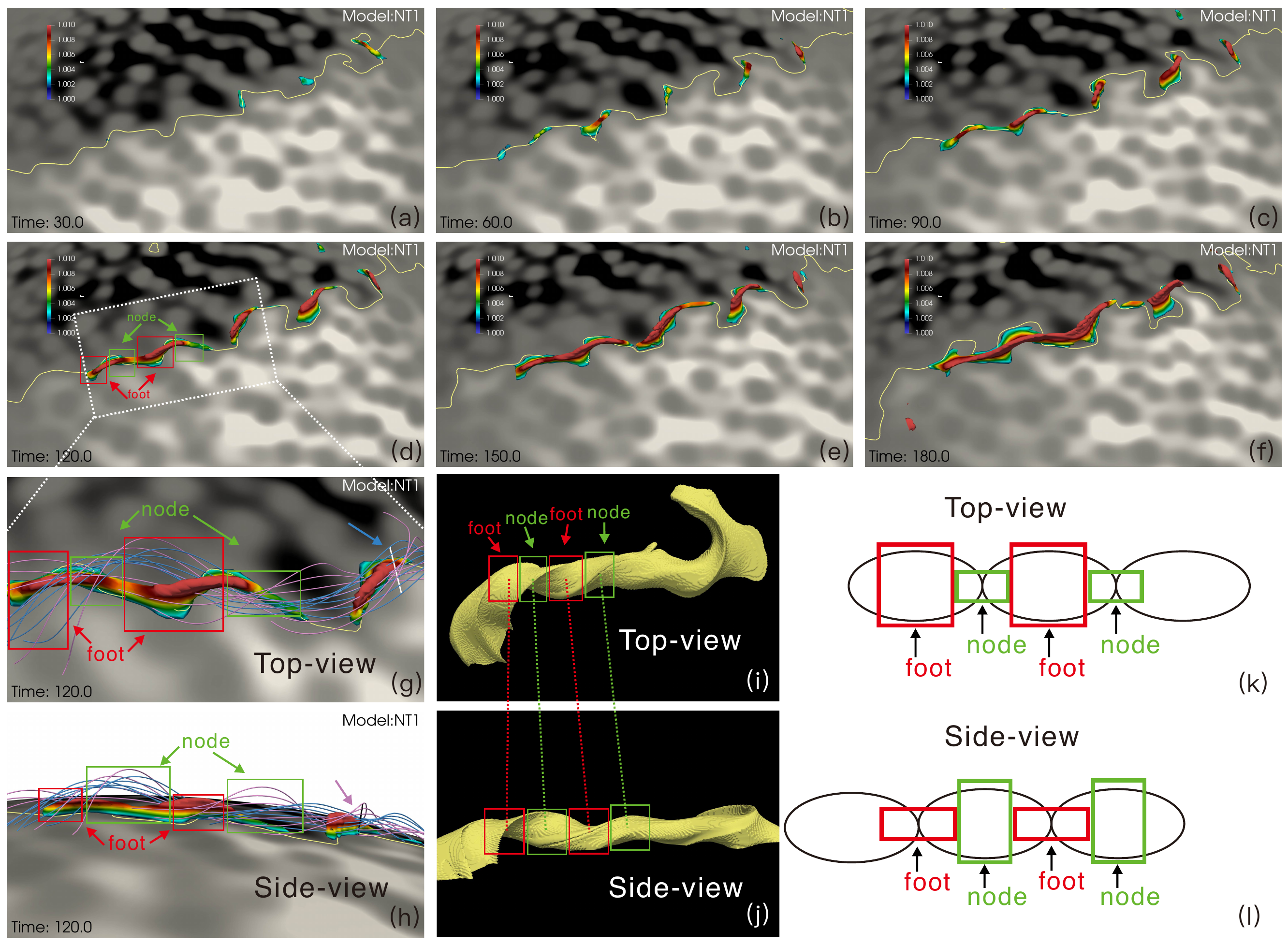}
\caption{Formation of magnetic dips and an MFR in model NT1 at time 30 (a), 60 (b), 90 (c), 120 (d), 150 (e), 180 (f). Magnetic dip regions are shown in the same way as in Figure~\ref{fig:NSdip}. Panel (g) and (h) show zoom-in views of the white dashed box of (d) with additional magnetic field lines from two viewing angles. The blue (horizontal line) and pink arrows indicate the horizontal and vertical lines to uniformly sample source points from which blue and pink magnetic field lines are integrated, respectively. Panels (i, j) show the isosurface of the twist number at 1.5 from two lines of sight, with the same ``foot" or ``node" substructure connected by dashed lines. Panels (k, l) depict cartoon diagrams of the cross-sectional area of MFRs from different views. The curve's width represents the cross-sectional area of the MFR as projected in the given line of sight. Red boxes and green boxes highlight ``foot" and ``node" substructures, respectively, in panels (d), (g), (h), (i), (j) and the cartoon (k, l).}
\label{fig:NT1dip}
\end{figure*}

Figure~\ref{fig:NT1dip} demonstrates magnetic dips along the PIL in the model NT1. At time 30 (Figure~\ref{fig:NT1dip} (a)), three small lumps of magnetic dips (LMDs) appear in the northwest half of the PIL. More lumps appear in the southeast half at time 60 (Figure~\ref{fig:NT1dip} (b)). These LMDs grow longer and start to connect to each other at later times (Figure~\ref{fig:NT1dip} (c) (d)). At time 150 and time 180 (Figure~\ref{fig:NT1dip} (e) (f)), the LMDs extend and connect to tall and long sheets of dip regions. Comparing the same moments in model NS, model NT1 has earlier formation and a faster growth rate of the magnetic dip regions. At time 120, we find a non-axisymmetric MFR hosting the dip sheet. The red boxes highlight the tall and old parts of the dip sheet, named ``foot", and between them, the green boxes show the newly-formed joint parts of dips, named ``node", in Figure~\ref{fig:NT1dip} (g-l). The dip regions at nodes have lower tops than the feet do. Figure~\ref{fig:NT1dip} (g) presents the magnetic field lines to clearly show that the MFR has a larger horizontal size at feet and a smaller size at nodes in the top view. Figure~\ref{fig:NT1dip} (h) with a lower side view shows the relative positions of the dip regions and the helical field lines of the MFR. The dip regions are at the lower part of the MFR below its axis and extend to the photosphere at feet. Using the code by \citet{Liu2016}, we calculate the twist number, plot the isosurface of 1.5 twist at time 120 in Figure~\ref{fig:NT1dip} (i) and (j), and find that the shape of the MFR resembles a screwed rope. We plot two cartoons in Figure~\ref{fig:NT1dip} (k) (l) to illustrate that 
the MFR has a ``foot-node-foot" variation along the axis from a top or side view. 
The locations corresponding to ``foot" (``node") have sheets of MFR field lines lying down (standing up), indicated by red (green) boxes. This pattern is also found in other moments and models. 

\begin{figure*}[htbp]
\centering
\includegraphics[width=\textwidth]{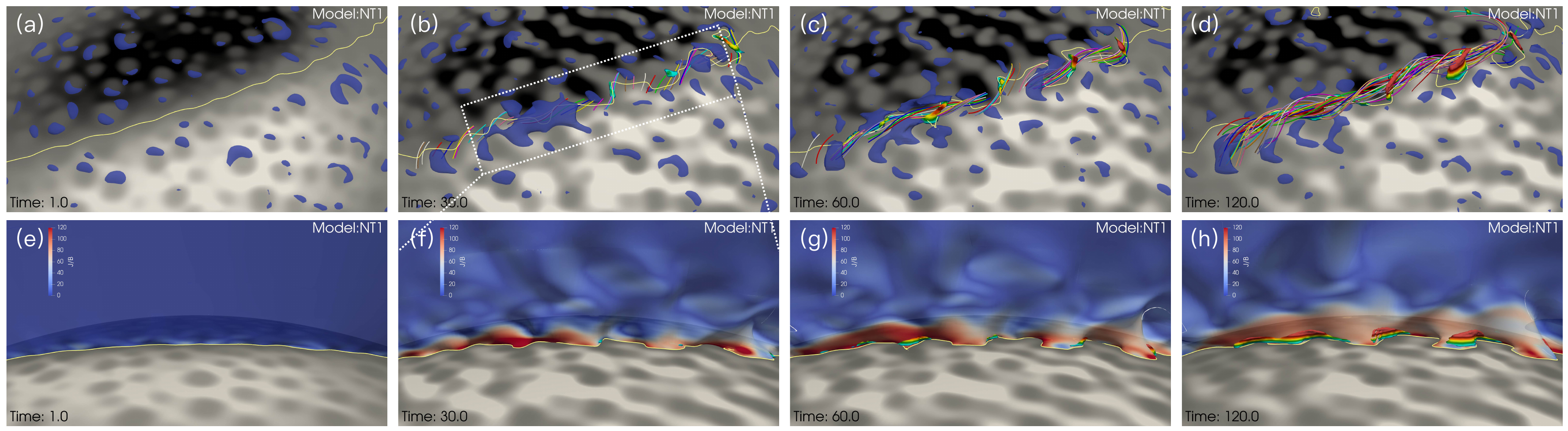}
\caption{Evolution of the ratio between the current density and the magnetic field intensity in model NT1 at time 1 (a, e), 30 (b, f), 60 (c, g), 120 (d, h). The blue translucent lumps in panels (a-d) are the isosurfaces for $J/B = 120$ with magnetic field lines and dip regions shown as figures above. In panel (e-h), the isosurfaces of $B_r = 0$ are plotted above the yellow PILs with blue-red colors showing $J/B$. The white dotted box in panel (b) indicates the field of side-view of panels (e-h).}
\label{fig:JB}
\end{figure*}

Figure~\ref{fig:JB} shows the ratio between current density and magnetic field intensity in model NT1. High values of the ratio can help to find twisted magnetic structures or current sheets. At time 1 as shown in Figure~\ref{fig:JB} (a), twisted magnetic structures only distribute above supergranular boundaries away from the PIL. As a result of the helicity condensation, the large $J/B$ regions appear above the PIL corresponding to shear arcades (Figure~\ref{fig:JB} (b)). The large $J/B$ contours retreat from newly formed magnetic dips (Figure~\ref{fig:JB} (c)), and concentrate around footpoints of the large MFR and overlying shear arcades (Figure~\ref{fig:JB} (d)). When looking at the isosurfaces of $B_r = 0$ from a side view (Figure~\ref{fig:JB} (e)-(h)), the $J/B$ values near the solar surface increase and form a discrete enhancement pattern in the early stage (Figure~\ref{fig:JB} (e) (f)). Each enhancement region of $J/B$ corresponds to a bundle of strongly sheared loops. After the formation of MFRs (Figure~\ref{fig:JB} (h)), the local intensity of $J/B$ decreases. 

\begin{figure*}[htbp]
\centering
\includegraphics[width=\textwidth]{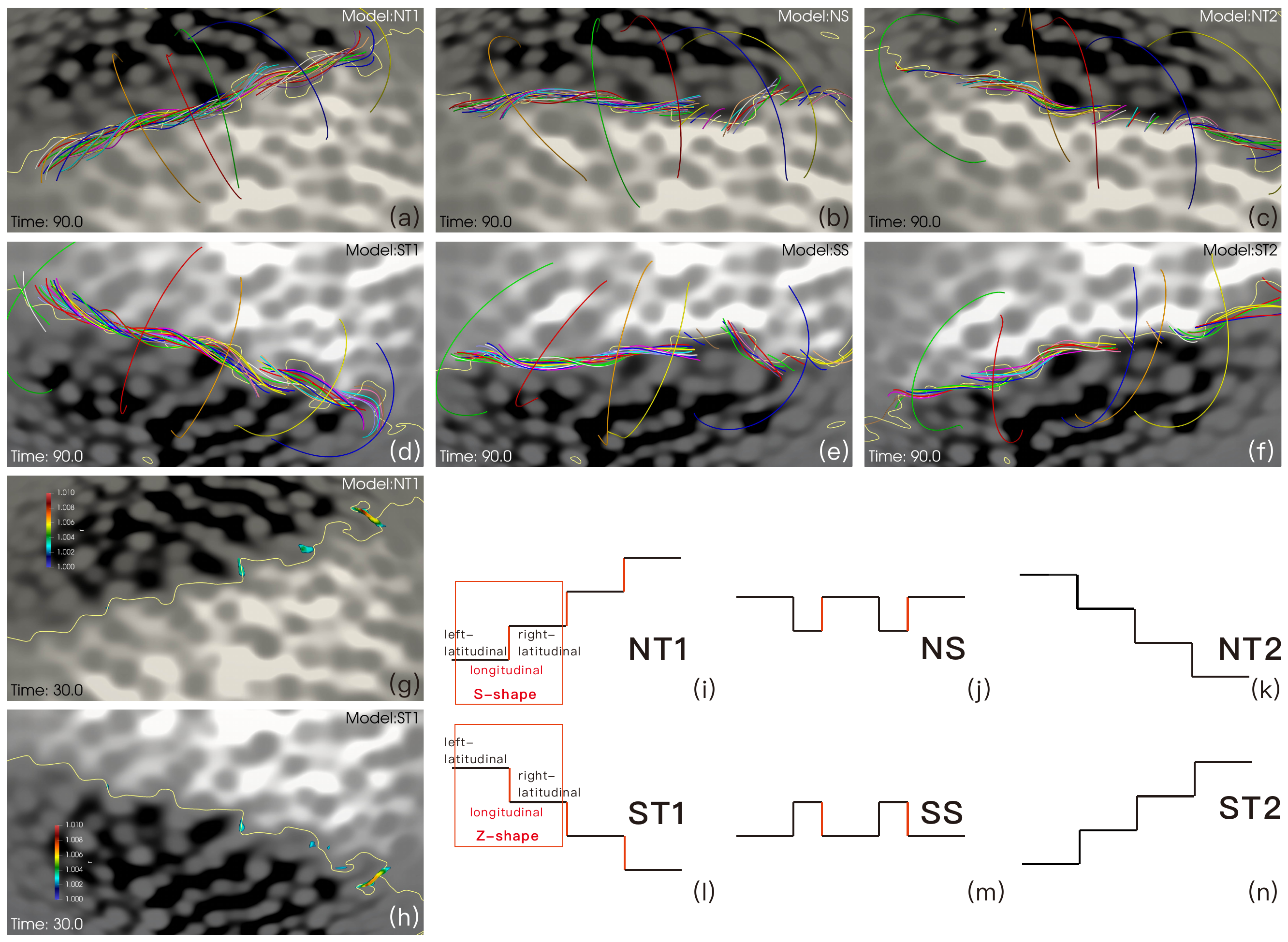}
\caption{Comparison of magnetic structures between the model NT1 (a), NS (b), NT2 (c), ST1 (d), SS (e), and ST2 (f) with different PIL tilt angles at time 90, and zigzag PIL patterns shown in model NT1 (g) and ST1 (h) at time 30 and summarized in the cartoon diagrams (i)-(n), in which the red longitudinal segments are favorable places for the formation of MFRs and dips. In the boxes in Figures (i) and (l), the S-shape and Z-shape local bending PILs are circled and can be divided into the left-latitudinal part, the longitudinal part, and the right-latitudinal part. The shape of the PIL at time 30 is very similar to the simplified diagram (g) (h), showing the ``up step" (S-shape) and ``down step" (Z-shape).}
\label{fig:NSTMFR}
\end{figure*}

\begin{figure*}[htbp]
\centering
\includegraphics[width=\textwidth]{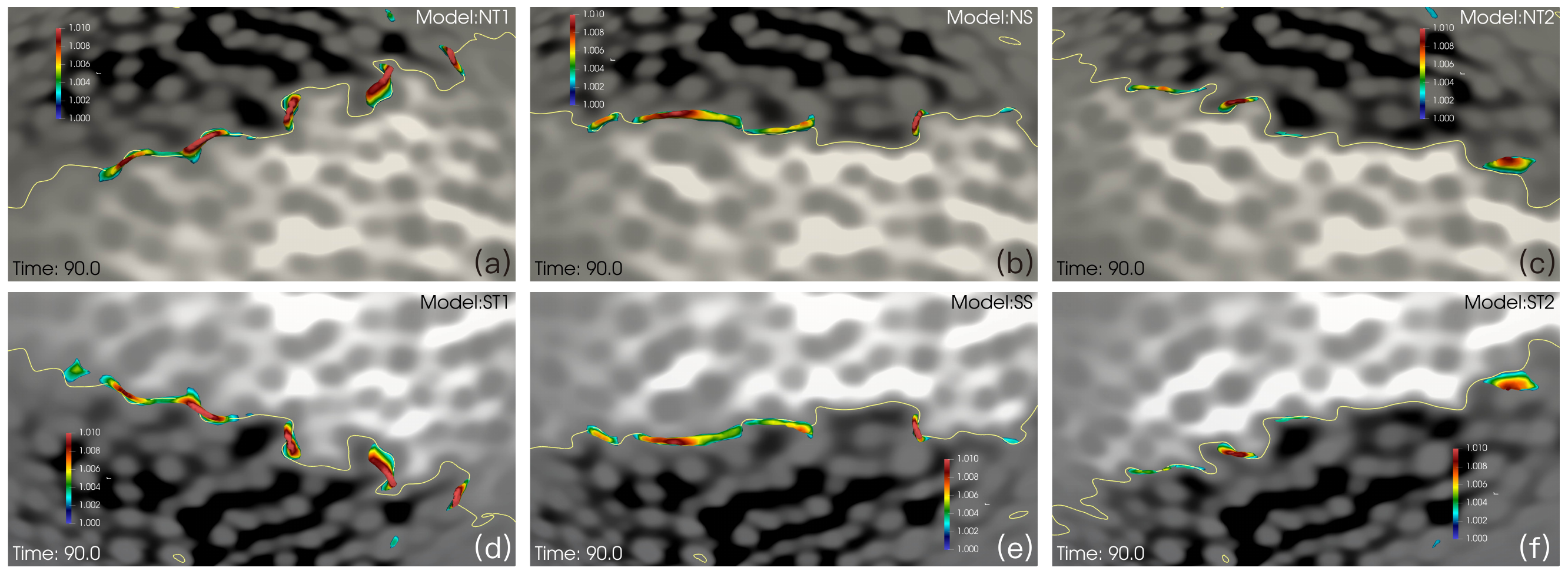}
\caption{Comparison of magnetic dips between the model NT1 (a), NS (b), NT2 (c), ST1 (d), SS (e), and ST2 (f) at time 90.}
\label{fig:NSTdip}
\end{figure*}

The PIL of the model NT1 (NT2) is initially set to the southeast--northwest (northeast--southwest) direction in the northern hemisphere. The PIL of model ST1 (ST2) is initially set to the northeast--southwest (southeast--northwest) direction in the southern hemisphere. From the comparison of magnetic field lines of six models in Figure~\ref{fig:NSTMFR} (a)-(f) at the same time 90, we find that the southeast--northwest PIL (northeast--southwest) style promotes the formation of MFRs in the northern (southern) hemisphere, and the northeast--southwest (southeast--northwest) PIL style suppresses the development of MFRs in the northern (southern) hemisphere since the MFRs in model NT1 and ST1 are longer and fatter than the standard models and the ones in models NT2 and ST2 are shorter and thinner than the standard models. In the case where the tilt angle of PIL is negative, i.e., the western end of the PIL is closer to the polar region, it is conducive to the formation of MFRs. The magnetic dip regions corresponding to the field line structures in Figure~\ref{fig:NSTMFR} (a)-(f) are shown in Figure~\ref{fig:NSTdip} (a)-(f). Model NT1 and ST1 clearly have more dip regions than model NT2 and ST2 (We counted the number of cells in the dip region for different models under the same selection criteria at time 90. The grid cells for the dip regions of models NT1, NS, and NT2 are 8193, 4460, and 2940, respectively. The dips in model NS, due to the relatively fewer bendings in the PIL, are connected together. Its length and width are much smaller than those in model NT1.), which indicates that PCFs preferentially form along southeast--northwest (northeast--southwest) tilted PIL in the northern (southern) hemisphere in agreement with statistical observations that PCFs have negative tilt angles \citep{Tlatov2016,Diercke_2019}. 
In these models, the magnetic structures in the northern and southern hemispheres have perfect mirror symmetry about the solar equator as a result of the symmetric initial condition and supergranular flows.

\section{Bending of PILs}
\begin{figure*}[htbp]
\centering
\includegraphics[width=\textwidth]{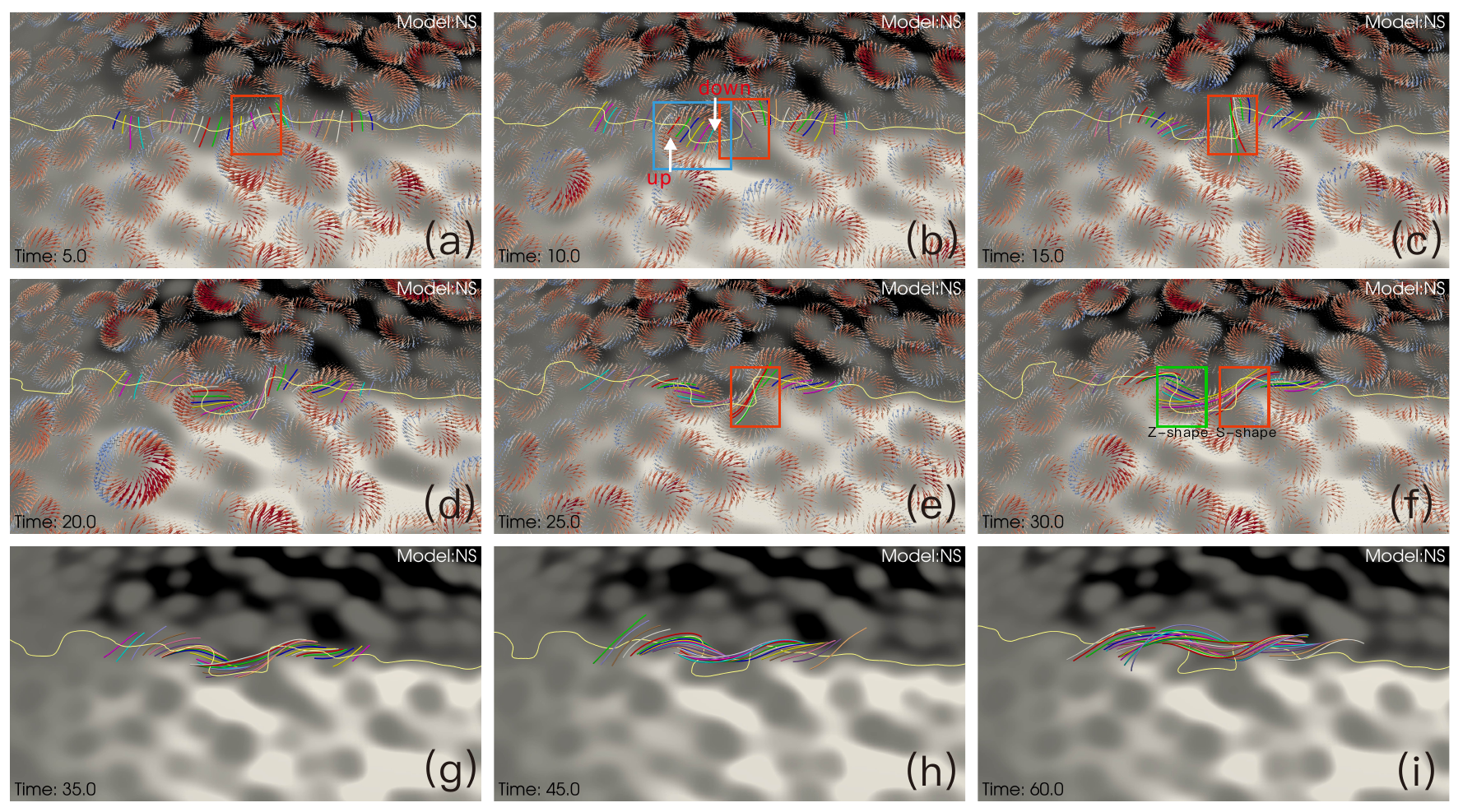}
\caption{Low-lying magnetic field lines above the bending PIL of model NS at time 5 (a), 10 (b), 15 (c), 20 (d), 25 (e), 30 (f), 35 (g), 45 (h), 60 (i). All magnetic field lines are integrated from sampled points of the PIL on a spherical surface with a radius of 1.0012. The horizontal supergranular velocities are shown in arrows colored by the divergence of the velocity field in blue-to-red colors and sized by the speed in panels (a-f). Two boxes respectively highlight the Z-shape (green box, weak shearing in the northern hemisphere) and S-shape (red box, strong shearing in the northern hemisphere) bending types of the PIL in (f). The blue box is used to indicate a single supergranular region where it is obvious that there are supergranular flows affecting the PIL shape. The white arrows indicate the direction (``up" or ``down") of PIL deformation due to the drive of the horizontal flows. The red boxes highlight bending PIL segments that are favorable for the shearing of magnetic field lines.}

\label{fig:PIL}
\end{figure*}

In all our models, the initially smooth PILs become sinuous under the varying driving motions of tens of supergranules nearby, in agreement with observations. We find some interesting bending patterns of PILs associated with MFR formation. The magnetic dips and small MFRs preferentially form at the longitudinal part of S-shape (Z-shape) PIL segments in the dextral (sinistral) MFRs, as shown in Figure~\ref{fig:NSTdip} and sketched in Figure~\ref{fig:NSTMFR} (i)-(n). Both S-shape and Z-shape PIL segments can be divided into the left-latitudinal part, the longitudinal part, and the right-latitudinal part. The S-shape (Z-shape) PIL presents an ``up step" (``down step") image from left to right, with the left-latitudinal part lower (higher) than the right-latitudinal part (Figure~\ref{fig:NSTMFR} (i) (l)). A Z-shape segment usually connects with an S-shape segment to form an extruding part towards the equator (Figure~\ref{fig:PIL} (f)).

To investigate why MFRs are preferentially formed in these regions, we analyzed the magnetic field variations around the bending PIL (Figure~\ref{fig:PIL}). Along the longitudinal part of the Z-shape PILs in model NS, the magnetic field lines above the longitudinal part are sinistral at the early time due to the fast bending of the PIL and slow spin of magnetic loops (Figure~\ref{fig:PIL} (a)-(c)). With the accumulation of negative magnetic helicity, the magnetic loops gradually spin to normal (Figure~\ref{fig:PIL} (e)) and then to dextral (Figure~\ref{fig:PIL} (f)) using the longitudinal part as a reference. Along the longitudinal part of the S-shape PILs, the PIL bending facilitates the dextral skew of magnetic loops, making them strongly sheared (Figure~\ref{fig:PIL} (a) (b)) and even aligned with the PIL (Figure~\ref{fig:PIL} (c)) at early times. The magnetic field lines in both the S-shape and Z-shape regions undergo clockwise rotation. However, due to the bendings in the PIL, the angle between the magnetic field lines and the locally bent PIL has changed. The accumulation of helicity along different path segments of the PIL results in varying local skew angles. When the S-shape PILs are fully developed, the elongated magnetic loops above spin clockwise further (Figure~\ref{fig:PIL} (d)-(f)), and develop the first magnetic dip region of a small MFR at time 30 (Figure~\ref{fig:NSdip} (a)). At the same moments, the magnetic loops above the S-shape PILs have larger skew angles than the magnetic loops above Z-shape PILs. Later, SMAs above the Z-shape PIL fuse into an MFR above the S-shape PIL to form a long MFR over the whole 
PIL segment (Figure~\ref{fig:PIL} (g)-(i)).

In early times, the skew angles of the short magnetic loops were strongly influenced by the direction changing of local PIL segments, which is probably related to the initial setting of a straight PIL and the supergranular motions along the PIL. The skew angles of the magnetic loops above the bending PILs, e.g., Z-shape or S-shape PILs, are very different from those of the loops above roughly straight PILs (Figure~\ref{fig:PIL} (b)-(g)). In all these models, although MFRs do not always stay above a bending PIL, any newborn MFR always first appears above a bending PIL (Figure~\ref{fig:NSTMFR} (a)-(f) and Figure~\ref{fig:NSTdip}). 

\begin{figure*}[htbp]
\centering
\includegraphics[width=\textwidth]{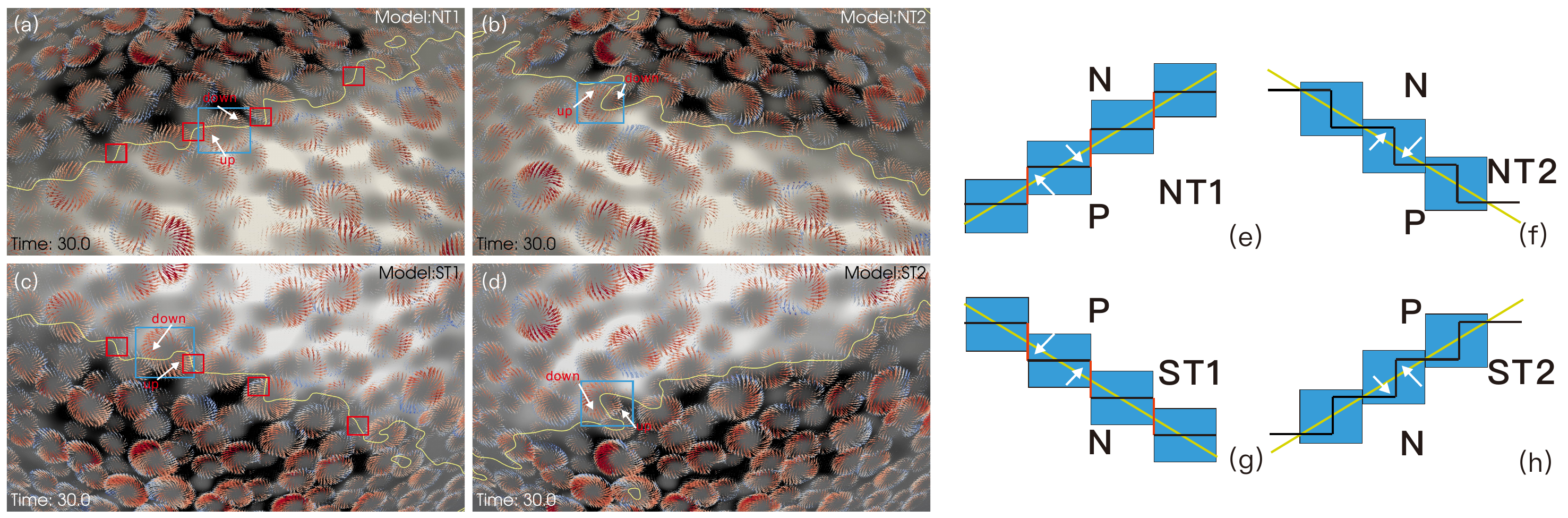}
\caption{Supergranular flows and PILs in model NT1 (a), NT2 (b), ST1 (c) and ST2 (d) at time 30, plotted in the same way as Figure~\ref{fig:PIL} (b). Panels (e)-(h) are cartoons illustrating the explanation of PIL deformation caused by supergranular flows in these models. The letter ``P" represents positive magnetic flux, while ``N" represents negative flux. Blue squares represent multiple supergranules, while yellow lines represent the initial smooth PILs. White arrows indicate the direction of deformation caused by the vortical motion of the supergranules. The black line with red segments, where MFRs preferentially form, represents the PIL after deformation.}
\label{fig:Vve}
\end{figure*}

The supergranular flows drive the movement of magnetic flux, which dominates the deformation and bending of the PIL. The scale of the extruding PIL is found to be similar to that of the supergranules. Figure~\ref{fig:PIL} (a)-(f) shows the snapshots for model NS, where the photospheric velocity field shown by arrows is overlaid on the magnetogram. It can be noticed that the velocity magnitude of the flows varies along the PIL in different supergranules marked by boxes. For the PIL segments that are deformed by the supergranular flows, there are two kinds of relative positions between the PIL and supergranular cells. When the PIL passes through the center of a supergranule (see the blue box in Figure~\ref{fig:PIL} (b)), the magnetic flux in the eastern (western) part of the supergranule is swept northward (southward) as the diverging supergranular flows clockwise rotate due to the Coriolis effect in model NS. Therefore the PIL within the supergranule tends to become higher on the left and lower on the right. The white arrows point out the tendency of the strong flow to drive the PIL to deformation. Where the PIL passes through the north (south) part of the supergranules, the PIL segment is pushed northward (southward) by the supergranular flows. The bending of the PIL becomes more zigzag where the PIL passes through the centers of the supergranules due to vortices. 

It seems that the northeast--southwest tilted PILs (NT2, ST1) have inapparent S-shape PIL segments with short longitudinal extension and apparent Z-shape PIL segments with long longitudinal extension. The shape of the PIL in the model ST1 at time 30, as shown in Figure~\ref{fig:NSTMFR} (h), can be approximated as a ``step-down stairs" in Figure~\ref{fig:NSTMFR} (l). Similarly, the southeast--northwest PIL in the model NT1 at time 30 (Figure~\ref{fig:NSTMFR} (g)) has dominating S-shape PIL segments and an overall ``step-up stairs" shape (Figure~\ref{fig:NSTMFR} (i)). The dextral (sinistral) MFRs in the northern (southern) hemisphere form more successfully and earlier because the shear is enhanced above S-shape (Z-shape) PILs. The differential rotation also injects a small amount of correct-sign magnetic helicity into the corona when the PIL is tilted.

Why do PILs with different tilt directions have different types of bending patterns? We conjecture that the supergranular motions are the key factors. In the northern hemisphere, supergranular flows rotate clockwise, pushing the eastern part of a PIL segment upward and the western part downward in a supergranular scale (Figure~\ref{fig:Vve} (a) (b)), resulting in an S-shape PIL segment within two neighboring supergranules in the model NT1 (Figure~\ref{fig:Vve} (e)) or a Z-shape PIL segment within one supergranule in the model NT2 (Figure~\ref{fig:Vve} (f)). In the southern hemisphere, supergranular flows rotate anticlockwise, pushing the eastern part of a PIL segment downward and the western part upward in a supergranular scale (Figure~\ref{fig:Vve} (c) (d)), resulting in a Z-shape PIL segment within two neighboring supergranules in the model ST1 (Figure~\ref{fig:Vve} (g)) or a S-shape PIL segment within one supergranule in the model ST2 (Figure~\ref{fig:Vve} (h)). The influence of the supergranules within the blue boxes in Figure~\ref{fig:Vve} (a)-(d) is consistent with the indicative cartoons. Several red boxes highlight the S-shape (Z-shape) PIL, which appears at the boundaries of the supergranules, in the northern (southern) hemisphere (Figure~\ref{fig:Vve} (a) (c)). The S-shape (Z-shape) PILs in the dextral (sinistral) MFRs enhance the magnetic shearing and are distributed in regions of high-density magnetic flux. It is natural to speculate that these types of regions are conducive to the formation of MFRs. During the early stage when the MFRs have not yet taken shape, due to the inhomogeneity and the time dependence of the supergranular flows, some PIL segments that are bent may recover to their original smooth shape. In these models, magnetic reconnection occurs on several PIL segments at supergranular boundaries, producing multiple, small MFRs simultaneously. These small MFRs along the PIL then grow and connect into a large axis-coherent MFR. 

\section{Discussion and Summary} \label{sec:floats}

To understand the magnetic structure of PCFs, we have done a series of local spherical magnetofrictional simulations driven by photospheric surface flows including differential rotation, meridional circulation, and supergranulation via time-dependent Voronoi tesselation, as listed in Table~\ref{tab:model}. The converging motions at the boundaries of the supergranules deflected by the Coriolis force form vortices which inject negative (positive) magnetic helicity into the corona in the northern (southern) hemisphere. The injected helicity at supergranular small scale inversely cascades and condenses to large scale along the PIL appearing as sheared arcades, which is consistent with the helicity condensation theory \citep{Antiochos_2013}. Dominated by supergranular helicity injection and helicity condensation, our models can well reproduce the hemispheric helicity/chirality pattern of PCFs, namely, negative magnetic helicity (dextral chirality) in the northern hemisphere and positive magnetic helicity (sinistral chirality) in the southern hemisphere. The global magnetofrictional models of \citet{Yeates_2012}, rely on the magnetic helicity transported from active regions at low latitudes to high latitudes, to overcome the effect of differential rotation and produce the hemispheric chirality pattern at high latitudes only in the rising phase of a solar cycle. Our local models do not need the aid of the transported low-latitude active-region helicity, which originates from solar dynamo in the convection zone. The HHR of filament chirality was reproduced and explained in both the rising and declining phase of the solar cycle by the global models of \citet{Mackay_2018}, which applied a large-scale, temporally and spatially averaged, statistically approximate helicity injection--condensation method \citep{Mackay_2014} and could not resolve supergranulations due to the limited resolution with such a large spatial and temporal coverage. Our local models use resolved realistic supergranular flows to replace the statistically averaged helicity injection and the effective surface flux diffusion in their models, and confirm that the helicity condensation counters the opposite-sign helicity injected by differential rotation along east--west oriented PILs. When the vortical speeds of supergranulations decrease, the helicity injected by the differential rotation can gradually become dominating and prevent the formation of MFRs in the correct helicity along the east--west oriented PILs. This result is consistent with the existence of a lower limit of the supergranular vorticity of $5\times 10^{-6}$ s$^{-1}$ from \citet{Mackay_2018}.

\citet{Filippov2017} observed sinistral and dextral fragmented filaments above winding PILs (see Figure 1 in \citet{Filippov2017}), and the dextral (sinistral) filament fragments seem to appear above the S-shape (Z-shape) PIL segments, which are well reproduced in our models (Figure~\ref{fig:NSTdip}) as the filament fragments are represented by magnetic dip regions. In his local flux-rope models of linear force-free field, the parasitic polarities of the magnetic field play an important role in the deformation of the PIL and the distribution of filament segments. In our models, the complexity of parasitic polarities is not included, the PILs are bent by vortical supergranular flows, and magnetic dip regions preferentially formed above the S-shape (Z-shape) PIL segments, where the shearing and nonpotentiality are enhanced by the local bending of PILs, in the dextral (sinistral) MFRs. It can be concluded that in the dextral (sinistral) MFRs, the S-shape (Z-shape) PIL segments located at the boundaries of supergranular cells, favor the formation of MFR structures. Applying this rule, it is possible to predict the chirality of a fragmented filament by the correlation between the shape of the PIL and the distribution of the filament fragments.

Observations show that the many PCFs are not exactly along the east--west direction, but have small tilt angles. In statistical research, \cite{Tlatov2016} found that filaments in the polar regions have negative tilt angles with western ends of filaments closer to the poles, which manifests the popular southeast--northwest (northeast--southwest) orientation of PCFs in the northern (southern) hemisphere. The results of models NT1, NT2, ST1, and ST2 with PILs in different tilt angles indicate that MFRs of PCFs preferentially form along southeast--northwest (northeast--southwest) directed PILs in the northern (southern) hemisphere, which is consistent with the observations. The reason is that S-shape (Z-shape) PIL segments dominate in the southeast--northwest (northeast--southwest) PILs and the formation of S-shape (Z-shape) PILs enhance the shearing of correct helicity to be favorable for the formation of dextral (sinistral) MFRs in the northern (southern) hemisphere. 

In the course of simulations, a noteworthy revelation emerged: the cross-sectional area of MFRs exhibited a departure from the uniformity observed in prior studies. Instead, it manifested an irregular, "foot-node-foot" periodic distribution, akin to the distinctive pattern found in a lotus root. Moreover, an intriguing spatial correlation was identified between these non-uniform cross-sectional areas and the distribution of magnetic dips when viewed from different perspectives.

The process of MFR formation can be summarized as follows: The bend of PILs into S-shape/Z-shape segments is primarily caused by horizontal supergranular flows. In the northern (southern) hemisphere, strong shearing near S-shape (Z-shape) PILs promotes the formation of small-scale MFRs. These initial small ropes gradually connect and merge into larger MFRs, where their positions within the larger ropes correspond to broad dips, resembling feet. PILs oriented in the southeast--northwest (northeast--southwest) direction are mainly influenced by the S-shape (Z-shape) PILs. This influence contributes to filaments aligning along PILs with a negative tilt at higher latitudes, although it's essential to note that this is just one contributing factor.

Some conclusions can be summarized as follows:

\begin{enumerate}
\item The converging vortical motions at the boundary of several supergranules, induced by the Coriolis force, inject the high-latitude helicity with the correct sign in the HHR observed in PCFs. 

\item The cross-sectional area of MFRs displays unevenness, and there is a spatial correlation between the distribution of MFR field lines and the shape of magnetic dip regions when observed from different views.

\item The bending of PILs caused by supergranular flows, forming S-shaped (Z-shaped) PIL segments, promotes the formation of dextral (sinistral) MFRs. The magnetic field supporting fragmented filaments may not necessarily require parasitic magnetic flux.

\item Experimental results with PILs of different tilt directions indicate a preference for the formation of PCFs along PILs with the western end close to the polar region.

\end{enumerate}

The magnetic models we got can serve as a start for investigating the plasma formation and eruption of quiescent filaments. We further evolve the MFR to a highly mature stage. The MFRs can rise to higher regions and find magnetic dip structures connected to the photosphere like feet. We will perform MHD simulations to explore the structure of quiescent filaments more realistically. These relevant works are currently underway and will be reported in follow-up papers. 

\acknowledgments

This research was supported by the Basic Research Program of Yunnan Province (202001AW070011), the National Natural Science Foundation of China (12073022, 11803031), the Strategic Priority Research Program of the Chinese Academy of Sciences (XDB0560000), the Key Research Program of Frontier Sciences, CAS (grant No. ZDBS-LY-SLH013), the Yunnan University Graduate Student Research and Innovation Foundation (KC-23233895), and Yunnan Key Laboratory of Solar Physics and Space Science under the number 202205AG070009. The numerical simulations were conducted on the Yunnan University Astronomy Supercomputer.

\software{MPI-AMRVAC \citep{Xia_2018}, 
          PDFI\_SS \citep{Fisher_2020}}

\clearpage


\clearpage
\bibliography{reference}{}
\bibliographystyle{aasjournal}


\end{document}